\documentclass{nature2}
\usepackage{graphicx}
\usepackage[usenames]{color}
\usepackage{xcolor}
\usepackage[export]{adjustbox}
\usepackage{amsmath}
\usepackage{amssymb}
\usepackage{chemformula,array}
\usepackage[symbol]{footmisc}

\title{Non-equilibrium Ion Transport in a Hybrid Battery Material}

\author{John Cattermull,$^{1,2,3}$ Ben Jagger,$^{2}$ Simon J. Cassidy,$^{1}$ Shobhan Dhir,$^{2}$ Phoebe K. Allan,$^{4}$ Mauro Pasta,$^{2\ast}$ Andrew L. Goodwin$^{1\ast}$}

\begin{document}

\maketitle

\begin{affiliations}
	\item Inorganic Chemistry Laboratory, Department of Chemistry, University of Oxford, Oxford, UK
    \item Department of Materials, University of Oxford, Oxford, UK
    \item Current Address: Department of Materials Science and Engineering, Stanford University, Stanford, CA, USA
    \item School of Chemistry, University of Birmingham, Birmingham, UK
\end{affiliations}

\begin{abstract}
Hybrid materials, which combine inorganic and molecular components, often exhibit structural flexibility that enables unusual functional responses. Among them, Prussian blue analogues (PBAs) are a promising class for post-lithium battery technologies. Here, we show that non-equilibrium transformation processes govern the charge-storage mechanism of a PBA electrode, K$_{\textbf 2}$Mn[Fe(CN)$_{\textbf 6}$]. Ostensibly, this behavior mirrors that observed in high-rate cycling of conventional cathodes such as LiFePO$_{\textbf 4}$, yet arises here for fundamentally different reasons—namely, low elastic moduli and cooperative distortions inherent to the hybrid framework. Using \emph{operando} methods, we show that framework flexibility limits transport kinetics and promotes collective, metastable pathways. Our results highlight new directions for PBA cathode optimisation, but also suggest a broader relevance of non-equilibrium mechanisms for mass transport in hybrid materials beyond PBAs alone.
\end{abstract}

Electrochemical processes, such as those occurring within electrode materials during battery cycling, often couple strongly to phase transformations.\cite{Bazant2013} The kinetics of such processes then involve a complex interplay amongst many competing factors, including charge carrier mobilities,\cite{Latz2011,Hatzell2020} elastic stress propagation rates,\cite{Cogswell2012} and the chemical potential gradients that develop at phase boundaries.\cite{Maier2004} A consequence of this complexity is the emergence of non-equilibrium thermodynamics in battery materials, whereby transformation mechanisms can vary with cycling rate and/or particle size. One particularly well-known example is that of LiFePO$_4$, which transforms \emph{via} a thermodynamically forbidden solid solution under high rates of cycling.\cite{Li2014,Liu2014,Lim2016} Another is that of Li(Ni,Mn,Co)O$_2$ (NMC), where the conventional wisdom of single-phase behaviour in Li compositions of $0.5<x<1$ was recently challenged by direct observation of phase separation using advanced \emph{operando} characterisation.\cite{Gent2016} In that system, state of charge heterogeneities arise due to composition-dependent reaction rates.\cite{Grenier2020,Park2021,Xu2022} The importance of understanding non-equilibrium behaviour is a crucial step to enable informed materials optimisation through the tuning of particle size, morphology, and disorder.\cite{Huang2001,Li2017,Lun2021} The relevance of these considerations to battery technologies based on hybrid chemistries remains largely unknown.\cite{Sada2023}

Prussian blue analogues (PBAs) are an increasingly prominent family of cathode materials for both Na- and K-ion batteries, yet one whose electrochemical transformation mechanisms are remarkably poorly understood.\cite{Dhir2020} From an application perspective, the key materials make use of cheap, abundant elements with the redox of Mn or Fe coupled to the reversible (de)insertion of Na- or K-ions; highly relevant criteria for next-generation battery materials to ease the dependence on critical minerals.\cite{IEA} The broader family of PBAs includes a large variety of systems.\cite{Lu2012,Pasta2014,Hurlbutt2018} Common to all is an open framework structure composed of molecular M$^\prime$-CN-M linkages, which impart a flexibility not observed in conventional oxide or polyanion cathode materials.\cite{Wessells2011b} More generally, the wider range of bonding interactions in hybrid inorganic materials is crucial to their remarkable performance in various applications, including perovskite optoelectronics.\cite{Li2017a} The PBA structure is made up of a simple cubic lattice decorated by alternating M$^{m+}$ and [M$^\prime$(CN)$_6$]$^{n-}$ ions with Na- or K-ions occupying the pore-space within. [M$^\prime$(CN)$_6$]$^{n-}$ vacancies (up to 1/3 per formula unit) are a common feature of PBA chemistry; these dictate the number of ions incorporated within the cubic framework cavities to balance charge. It is the softness of the molecular interactions in PBAs that means the materials can be synthesised at room temperature from aqueous solution;\cite{Sharpe1976,Wu2016} but this softness also promotes a variety of low-energy distortions.\cite{Moritomo2011,Cattermull2021} 

It has become increasingly apparent that the structural distortions in PBAs are intimately connected to their electrochemistry.\cite{Jiang2019,Cattermull2022} Structural distortions are a particular characteristic of low-vacancy PBAs, due to increased connectivity and higher alkali-metal cation content.\cite{Cattermull2023a} Such low-vacancy PBAs promise higher theoretical capacities, but cycle \emph{via} multiple phases and so have a reduced rate capability.\cite{Bie2017,Fiore2020} In the pristine state of low-vacancy PBAs, distortions are dominated by cooperative displacements of the alkali-metal cation,\cite{Cattermull2023} whereas when the cathode is fully charged distortions arise from framework strain.\cite{Bie2017} By contrast, high-vacancy PBAs show no cooperative distortions: their disordered cubic structure cycles \emph{via} a solid solution with high rate capability.\cite{Wessells2011} From a design perspective, the suppression of distortions in low-vacancy PBAs has been an obvious target, since it would allow high rate capability whilst preserving high capacities.\cite{Jiang2019} However, without a mechanistic understanding of how the phase transformations of low-vacancy PBAs couple to the electrochemistry, it is difficult to strategically target higher performing PBA electrode chemistries.\footnote[2]{Simulating PBAs \emph{ab initio} is notoriously challenging due to large unit cells, high degrees of vacancies and disorder, as well as a landscape of low energy distortions shown to fluctuate with moderate changes in temperature.\cite{Hosaka2021,Deng2021}}
Careful experimental study of these mechanisms is therefore key to unravelling the complex electrochemical behaviour and its interplay with composition and structure.

To address this fundamental problem, we study here the K-ion cathode, K$_2$Mn[Fe(CN)$_6$], as a model system. Our choice of material was influenced by the following considerations. First, it is this PBA that can be prepared as an anhydrous, vacancy-free material, with well-separated monolithic particles.\cite{Dhir2024} \footnote[3]{Na-ion analogue the structure collapses by 20\% in volume while the PBA framework.\cite{Wang2015,Brant2019}} This is important if we are to investigate electrochemical cycling effects independent from complications of varied particle morphology and water-based high-voltage degradation. Second, amongst PBAs its structure and associated framework distortions are particularly well-characterised.\cite{Cattermull2021,Cattermull2023} Third, K$_2$Mn[Fe(CN)$_6$] has a high theoretical capacity of 155\,mA\,h\,g$^{-1}$ due to low-vacancy content which, combined with the high operating voltage of the Mn$^{3+}$/Mn$^{2+}$ couple (close to 4\,V vs. K$^{+}$/K), enables K-ion batteries with a specific energy density competitive with state-of-the-art LiFeO$_4$/graphite cells.\cite{Dhir2024} We use \emph{operando} X-ray absorption spectroscopy (XAS) and X-ray diffraction (XRD) to study the phase transformation mechanisms of this PBA electrode during a single charge/discharge cycle. We will come to show that its electrochemistry is governed by multiphase behaviour due to subtle structural changes, and that the structural transformations that occur are non-equilibrium in their nature. The underlying cause of this non-equilibrium behaviour is the flexiblity of the PBA framework, which acts to couple ion transport with framework deformations and inhibits domain wall motion. These considerations are widespread amongst hybrid materials, and hence our results are likely to be relevant to challenges of ion transport in the broader family of hybrid systems.

\section*{Charge behaviour}
\begin{figure}
	\centering
	\includegraphics{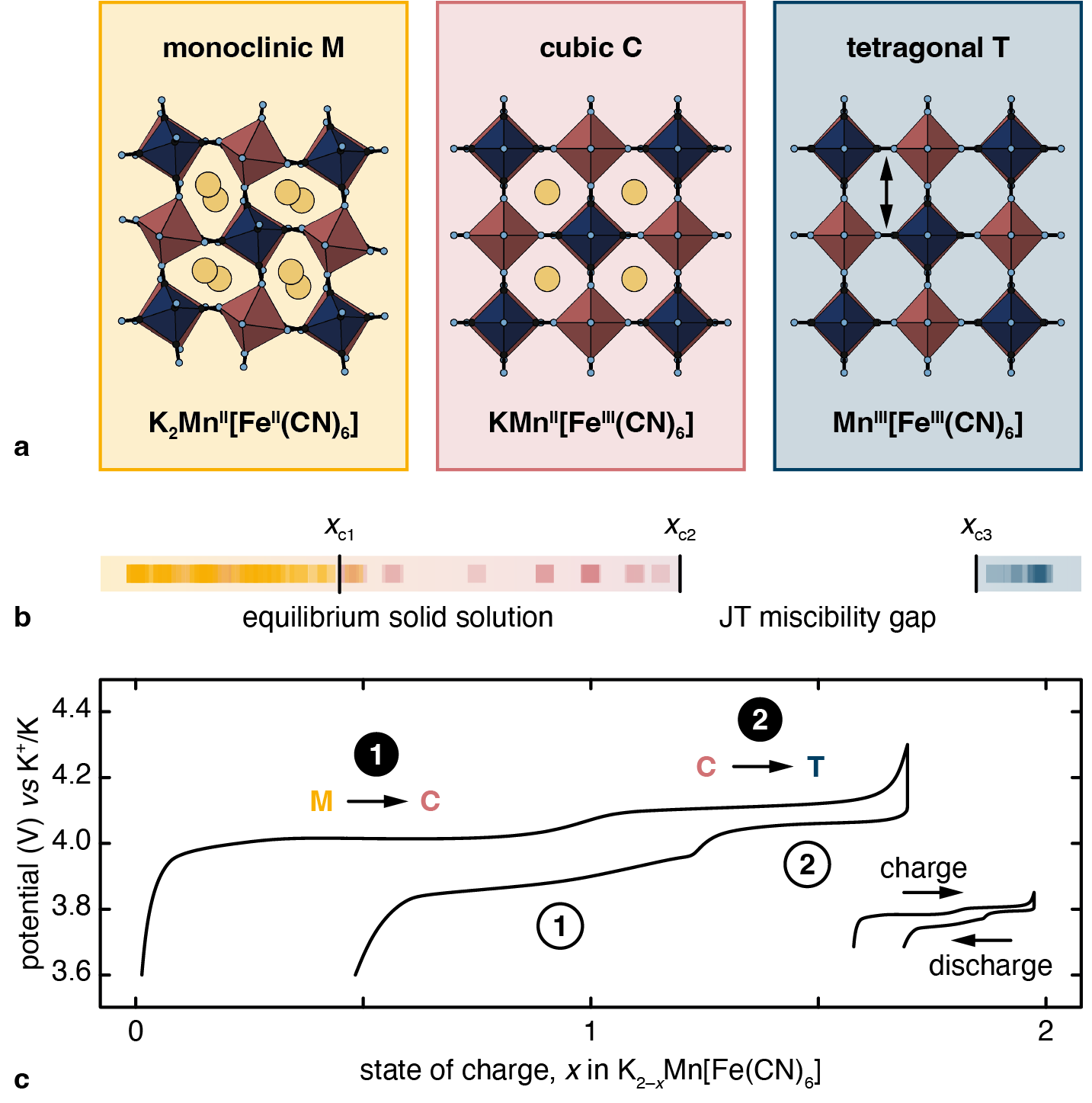}
	\caption{\footnotesize Structures of K$_{2-x}$Mn[Fe(CN)$_6$] \emph{via} chemical and electrochemical preparation. (a) Schematic 2-dimensional representation of the three PBA phases present during electrochemical cycling. In the JT distorted Mn[Fe(CN)$_6$] arrows included to represent the anisotropy due to JT-active Mn. K are shown as amber spheres, FeC$_6$ are shown as dark blue octahedra with black nodes and MnN$_6$ as brown octahedra with light blue nodes. (b) Samples of composition K$_{2-x}${\sf M}[{\sf M$^{\prime}$}(CN)$_6$]$_y$, which were reported in Refs.~\citenum{Cattermull2023a}, \citenum{Harbourne2024a}, and \citenum{Bostrom2022} are plotted on an axis of  $x$ and coloured according the unit cell symmetry they were assigned. A line at $x\sim 0.45$ is included as a guide to the eye for a threshold value, $x_{\rm c1}$. The apparent JT miscibility gap for intermediate composition of JT-active metal is similarly highlighted with $x_{\rm c2}$ and $x_{\rm c3}$ as the solubility limits of the cubic phase (1.15) and tetragonal phase (1.85), respectively. (c) The charge-discharge profile of the K$_2$Mn[Fe(CN)$_6$]/K half-cell. Regions \textbf{1} and \textbf{2} are highlighted represent the monoclinic to cubic transition and cubic to tetragonal transition, respectively and are discussed in the main text. }
	\label{fig1}
\end{figure}

On charging, the K$_2$Mn[Fe(CN)$_6$] cathode transforms \emph{via} two phase transitions, with the pristine and fully charged states having different structures, and each different again to a third partially-charged intermediate phase [Fig.~\ref{fig1}a].\cite{Bie2017} The pristine K$_2$Mn[Fe(CN)$_6$] has a monoclinic structure owing to the K-ion slide distortion---a cooperative off-centering of K$^+$ ions within the cavities toward an edge of the surrounding anionic PBA framework that couples to octahedral tilts.\cite{Cattermull2023} Under (equilibrium) chemical control, low-vacancy compositions such as ours can prepared with K-ion content as low as 1.6 per formula unit whilst still crystallising as a single-phase monoclinic structure.\cite{Jiang2017} For lower K-ion contents, these PBAs adopt a cubic structure instead, as a result of the weakened driving force for cooperative K-ion off-centering [Fig.~\ref{fig1}b].\cite{Cattermull2023a} The fully-charged K$_0$Mn[Fe(CN)$_6$] contains Jahn-Teller (JT) active Mn(III) which induces a tetragonal distortion of the framework through a cooperative JT effect. The associated strain is why JT solubility is so poor in the cubic phase of low-vacancy PBAs; for example, in the system Mn$^{\rm II}_x$Cu$^{\rm II}_{1-x}$[Pt], a miscibility gap is observed for $0.15<x<0.85$ [Fig.~\ref{fig1}b].\cite{Harbourne2024a} Hence---from an equilibrium perspective---one expects conversion between the three phases on charging K$_{2-x}$Mn[Fe(CN)$_6$] to occur first by phase change from monoclinic to cubic symmetries at some critical K-ion content $x\simeq0.45$, followed by a second phase change from a cubic phase with $x\simeq1.15$ to a tetragonal phase with $x\simeq1.85$. 

In practice, as K$_2$Mn[Fe(CN)$_6$] is cycled, current flows at two separate voltage plateaux, each contributing similar capacity, corresponding to the (de)insertion of one mole of K$^+$ ions from the PBA and an associated phase transition [Fig.~\ref{fig1}c].\cite{Bie2017} Referring to Fig.~\ref{fig1}c, in region \textbf{1} the pristine (K$_2$Mn[Fe(CN)$_6$]) phase is converted to the intermediate (K$_1$Mn[Fe(CN)$_6$]) phase almost instantly with charging, evidenced by the flat voltage plateau. The immediate nature of this transition is evidence of non-equilibrium behaviour, since chemical preparation of the PBA can produce range of compositions with K-ion concentrations between $0<x<1$, all single phase [Fig.~\ref{fig1}b]. In region \textbf{2}, the conversion of K$_1$Mn[Fe(CN)$_6$] $\rightarrow$ K$_0$Mn[Fe(CN)$_6$] is coupled with the emergence of a cooperative JT distortion from the presence of Mn$^{3+}$.  With this in mind it is expected that K$_{2-x}$Mn[Fe(CN)$_6$] should phase separate for intermediate values of $x$. Although these two transformations have been observed experimentally in previous studies,\cite{Bie2017,Jiang2019,Peng2019,Deng2021,LePham2023} little is known about the transformation mechanisms themselves. In particular, there is no quantitative understanding of the population of each phase for a given state of charge, nor that of any compositional or structural changes within the phases themselves. We anticipate that the mechanisms of the transformations will be central to rationalising the performance of the cathode material. 

\begin{figure}
	\centering
	\includegraphics{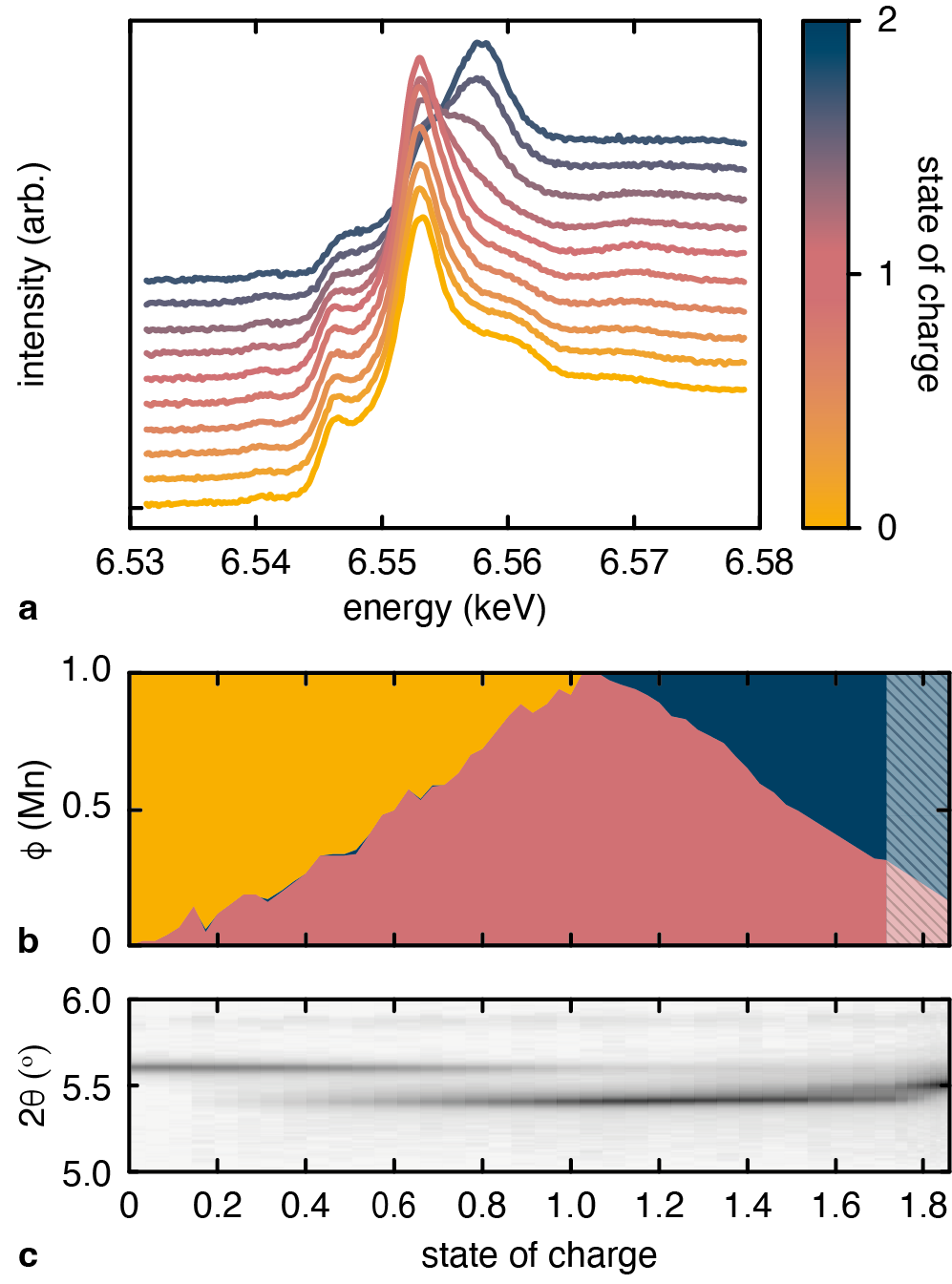}
	\caption{\footnotesize \emph{Operando} characterisation of the K$_{2-x}$Mn[Fe(CN)$_6$] cathode. (a) Normalised XAS profile for the Mn K-edge with selected curves offset vertically by a constant amount and coloured by state of charge. (b) MMF phase fractions for the fixed MMF refinement of the XAS data. The pristine spectrum is in amber, the intermediate spectrum in pink and the Mn(III) component in dark blue. The shaded region is extrapolated to match up with the XRD plot, where a constant voltage hold at 4.3\,V drove the state of charge to a minimum of $x=1.84$. (c) Film plot for region around the 200 XRD reflection of the parent cubic structure to show the structural phase changes on cycling. The monoclinic phase appears to monotonically convert to the larger cubic phase before finally converting to the smaller tetragonal phase late on charge.}
	\label{fig2}
\end{figure}

Our starting point in linking electrochemistry to structural transformations was to use \emph{operando} XAS to track the changes in oxidation state during cycling \emph{via} the K-edge energy profiles of Fe and Mn. The Fe K-edge shift turns out to be relatively insensitive to charge state as a consequence of covalency in the Fe--CN interaction. By contrast, the Mn K-edge profile tracks the oxidation state of the PBA and is even indirectly sensitive to Fe charge state [Fig.~\ref{fig2}a]. The clearly-defined voltage plateaux associated with the redox of each metal indicates that the XAS data ought to be separable into components of each charge state species---namely, pristine, fully charged, and a partially-charged intermediate. For that reason, we employed Metropolis matrix factorisation (MMF) to analyse the XAS spectra for Mn.\cite{Geddes2019} MMF refinement of the \emph{operando} XAS data using three components confirmed that the Fe$^{\rm 3+/2+}$ redox couple is active in the first charge plateau, and Mn$^{\rm 3+/2+}$ in the second charge plateau [Fig.~\ref{fig2}b]. The XAS/MMF analysis also enables accurate normalisation of capacity by state of charge, represented as units of $x$ in K$_{2-x}$Mn[Fe(CN)$_6$]. A more detailed discussion of the MMF fitting is given in the supplementary materials. Importantly, the link between capacity and state of charge now established allows normalisation of \emph{operando} XRD data collected under similar conditions [Fig. \ref{fig2}c].

\section*{First charge plateau}
To interpret our \emph{operando} XRD data, we turn our attention first to the charge region \textbf{1}, where the pristine phase begins to convert into an intermediate phase early on in the charging process [Fig.~\ref{fig2}c]. Refinement of the XRD patterns allows the phase fractions to be expressed as a function of state of charge, as summarised in Fig.~\ref{fig3}a (refinement method is discussed in the supplementary materials). We see a gradual coupled change in monoclinic and cubic phase fractions, that we will come to show indicates K-ions are not being removed homogeneously from the PBA---\emph{i.e.}  as a single bulk phase of composition K$_{2-x}$Mn[Fe(CN)$_6$] ($0<x<1$). From an equilibrium perspective, the pristine monoclinic phase is stable on depotassiation  until $x\sim0.45$ before its conversion to the intermediate cubic phase [Fig.~\ref{fig1}b] and hence one would expect onset of phase transformation only at the relevant charge state. The experimental departure from this picture suggests K-ion extraction is under kinetic control during electrochemical K-ion extraction, with a barrier to extraction that is dependent on loading, $x$. Our interpretation here is based on similar behaviour reported for Ni-rich NMC cathodes, where concentration-dependent diffusivity produces a ``core-shell'' structure of lithium-poor peripheries and lithium-rich cores in single-particle cathodes at the beginning of charge, giving rise to a closely-related \emph{operando} XRD profile.\cite{Xu2022}

\begin{figure}
	\centering
	\includegraphics{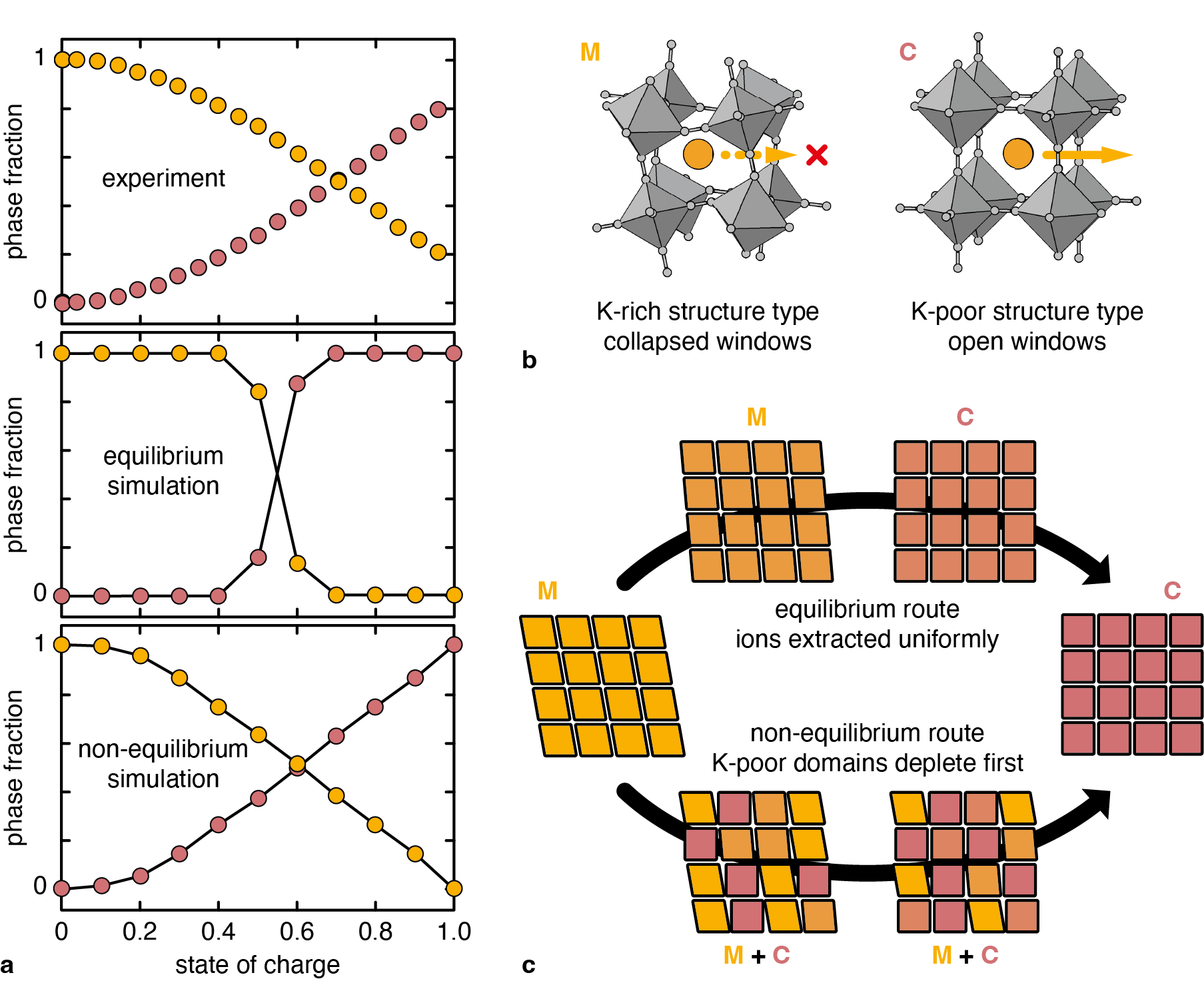}
	\caption{\footnotesize \emph{Operando} XRD analysis of K$_{2-x}$Mn[Fe(CN)$_6$] first charge plateau. (a) XRD refinement phase fractions as compared to the predicted phases from equilibrium and non-equilibrium simulations. (b) Schematic structural representation of the effect of opening the framework windows on K-ion transport when the monoclinic phase transforms into the cubic phase. (c) Evolution of coherent scattering domains with K-ion extraction in the case of equilibrium and non-equilibrium transformations. Based on the simulations, we interpret the experimental phase fraction evolution to represent the non-equilibrium route here.}
	\label{fig3}
\end{figure}

What would drive such a non-equilbrium pathway in PBAs? In the highly-potassiated state, the PBA framework is collapsed to maximise its interaction with the K$^+$ ions; the degree of collapse scales with K-ion concentration, pinning the K-ions less strongly as K-ion composition is reduced.\cite{Cattermull2021,Cattermull2023} The picture that emerges is one where K-ion mobility increases for higher values of $x$, meaning that as K-ions are removed from crystallites of K$_2$Mn[Fe(CN)$_6$], subsequent removal is favoured in locally depleted regions where the framework is opened up [Fig.~\ref{fig3}b]. A simple model that incorporates this composition-dependent mobility captures surprisingly well the key phase behaviour observed experimentally [Fig.~\ref{fig3}a] (see supplementary materials for further discussion). So our experimental data support a model in which phase transformation occurs heterogeneously throughout PBA crystallites [Fig.~\ref{fig3}c].

This behaviour in K$_2$Mn[Fe(CN)$_6$] contrasts with that of other better-studied cathode chemistries. In LiFePO$_4$, for example, there is a strong thermodynamic driving force for phase separation such that intermediate compositions Li$_x$FePO$_4$ are unstable.\cite{Padhi1997} Lithium mobilities in that system are clearly much higher in the absence of an applied potential than the potassium mobilities in PBAs, such that kinetically-obtained intermediate compositions Li$_x$FePO$_4$ are able to relax quickly to LiFePO$_4$/FePO$_4$ mixtures (\emph{e.g.}\ on a $\sim$10s timescale after charging at 10\,C).\cite{Liu2014} No such relaxation to single-phase equilibrium compositions is possible for our PBA samples on the experimental timescale we probe. So, despite its apparent simple biphasic plateau, K$_2$Mn[Fe(CN)$_6$] is actually more similar to Ni-rich NMC and almost in direct contrast with LiFePO$_4$, in that the combination of slow K-ion mobility, relatively weak electrostatics, and K-composition-dependent kinetics transforms what should be a solid-solution transformation into a two-phase process.

\section*{Second charge plateau}
\begin{figure}
	\centering
	\includegraphics{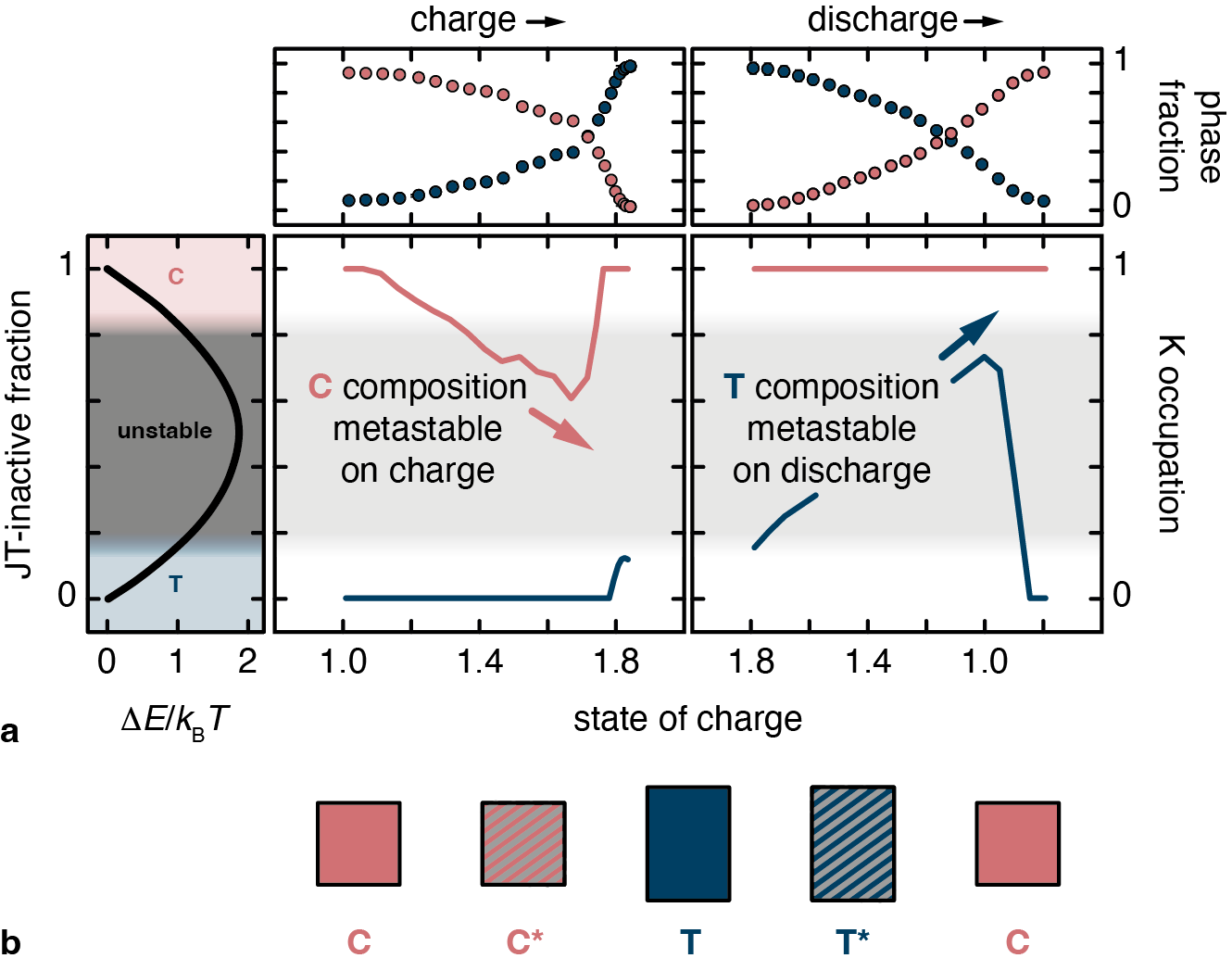}
	\caption{\footnotesize \emph{Operando} XRD analysis of the high voltage plateau. (a) XRD refinement phase fractions are plotted as a function of state of charge. Using the  strain map from Ref.~\citenum{Harbourne2024a}, the strain of the incumbent phase can be visualised based on necessary change in K-ion occupancy. (b) The strain present in the PBA is illustrated schematically. On charge, the strain in the intermediate phase, labelled C, builds up with little change in phase fraction before relief of that strain by accelerated phase transformation. On discharge, the reverse follows, the strain builds up in the fully charged phase, labelled T, with slow phase conversion before faster phase conversion relieves the strain}
	\label{fig4}
\end{figure}

We turn our attention now to the region \textbf{2}, where cubic K$_1$Mn[Fe(CN)$_6$] is converted into tetragonal K$_0$Mn[Fe(CN)$_6$] [Fig.~\ref{fig1}]. This composition range differs from the first in that, under equilibrium conditions, no intermediate phases have been isolated---a consequence of the strain-driven JT insolubility discussed above. Our XAS data showed a clear phase change relating to Mn, which was also visible in the \emph{operando} XRD patterns [Fig.~\ref{fig2}b,c]. Constrained refinements of the XRD traces allowed us to quantify the relative fractions of cubic and tetragonal phases, but we found the quantities of those phases were not sufficient to account for the known charge states [Fig.~\ref{fig4}]. For example, at a state of charge of $x=1.50$, only 20\% of the material has converted to the fully-charged phase. Consequently, the composition of the two phases present must itself be changing during charge; evidence for accommodation of Mn$^{3+}$ within the intermediate phase comes from an accompanying decrease in lattice parameter [Fig.~S9]. It is only near the very end of the charge plateau---and indeed on application of a final constant-voltage hold---that the fully-charged phase suddenly emerges in appreciable quantities [Fig.~\ref{fig4}]. On discharge, a clear hysteresis in phase fraction evolution is observed, with the fully-charged tetragonal phase now dominant much longer than anticipated. Taken together, these observations are consistent with a significant kinetic barrier to interconversion between the two phases [Fig.~\ref{fig4}]. The sluggish transformation kinetics mean that the dominant phase present must accommodate strain associated with K-ion concentration changes and accompanying Mn redox beyond its equilibrium stability field [Fig.~\ref{fig4}].

Conventional multiphase cathode materials rely on high elastic constants to drive the phase boundary wave on charge and discharge.\cite{Cogswell2012} However, the softer bonding interactions in the PBA framework and stronger elastic compliance allow strain to build up in the framework without inducing phase change, reducing the driving force for phase boundary wave propagation.\cite{Harbourne2024} Any delay in phase transformation fundamentally limits the rate capability of a material,\cite{Xu2017} which is likely why PBAs exhibit low charge storage capacities at high cycling rates.\cite{Deng2021} In other systems---\emph{e.g.} LiFePO$_4$---wave propagation can accommodate the formation of metastable solid solution phases when cycling rates are faster than phase transformation process. That is to say the coherency strain in ceramics with higher elastic constants aids diffusion,\cite{Bazant2013} but in more flexible materials such as PBAs, the flexibility hampers ion diffusion by slowing the phase transformation. A corollary of this argument is that the lack of phase transformations in high-vacancy PBAs therefore explains the dramatically superior rate capability found in those systems.\cite{Wessells2011}

\section*{Discussion}

Our key result is to have shown how non-equilibrium transformation mechanisms extend from conventional cathode materials to hybrid materials such as PBAs, albeit for different microscopic reasons. Nevertheless our study also has the practical outcome of identifying potential strategies for optimisation of electrochemical performance in K-ion PBAs. For example, improvement to the kinetics in the first charge step might be achieved by lowering the initial K-ion concentration\cite{Cattermull2021}, exchanging Mn for smaller transition-metals,\cite{Cattermull2023} or pinning open the framework with a low level of Cs$^+$ doping---all of which would stabilise the cubic phase to higher K-ion concentrations. These strategies would have to be tensioned against the associated reduction in specific energy density of the material in each case. The more conventional strategy of targeting smaller particle sizes\cite{Bruce2008} increases the relative number of available K-ion extraction sites, but does so whilst actually increasing the fraction of K-ions with high kinetic barriers to extraction.  This same strategy would be advantageous, however, for the second charge step, accelerating the phase transformation from cubic to tetragonal and potentially improving reversibility. An alternative consideration to varying particle size is that using different hexacyanometallate vacancy correlations\cite{Simonov2020} to optimise the interplay between vacancy fraction and phase stability---the ultimate goal being to stabilise the undistorted intermediate cubic phase for all K-ion compositions at as low a vacancy fraction as possible. These twin directions of controlling particle size and defect engineering will alter the non-equilibrium behaviour of PBAs and we intend to explore both aspects in follow-up studies.

The microscopic ingredients responsible for non-equilibrium ion transport in K$_2$Mn[Fe(CN)$_6$] are relevant to all PBAs, and to hybrid materials more generally. For example, distortion mechanisms and their coupling to A-site occupancies obey universal trends amongst PBAs,\cite{Bostrom2022,Cattermull2023a} and so will affect ion-transport mechanisms of other important PBA electrode materials (\emph{e.g.}\ Na$_2$Fe[Fe(CN)$_6$]).\cite{Wang2015} In metal--organic frameworks---which generally share the elastic compliance of PBAs\cite{Tan2011}---sluggish transformation kinetics associated with guest-driven phase transformations (\emph{e.g.}\ in MIL-53\cite{Liu2008} or DUT-49\cite{Krause2016}) might now be understood through the same lens used here to rationalise the second charge mechanism of K$_{2-x}$Mn[Fe(CN)$_6$]. Likewise in hybrid perovskite photovoltaics, where both ion diffusion and strain localisation are key to chemical stability and long carrier lifetimes, respectively.\cite{Guzelturk2021,Dubajic2025} Whatever the particular family, our results highlight that the unique and general features of hybrid materials---their diversity of structural distortions and anomalous mechanics---will be important design tools not only for equilibrium properties (now routinely explored\cite{Saparov2016,Bennett2017,Li2017a,Bostrom2018,Kronawitter2024}) but also for controlling the mechanism of functional response away from equilibrium.

\section*{Acknowledgments}
A.L.G. gratefully acknowledges C. P. Grey for useful discussions. We acknowledge the provision of beam time by Diamond Light Source, U.K. and the technical assistance of Veronica Celorrio on beamline B18 and Sarah Day on beamline I11. Funding: this work was generously supported by the European Research Council (advanced grant 788144) and the Henry Royce Institute (through U.K. Engineering and Physical Science Research Council grant EP/R010145/1). Competing interests: The authors declare no competing interests.

\clearpage

\section*{References}
\bibliography{K2MnFe}

\begin{thebibliography}{10}
\expandafter\ifx\csname url\endcsname\relax
  \def\url#1{\texttt{#1}}\fi
\expandafter\ifx\csname urlprefix\endcsname\relax\def\urlprefix{URL }\fi
\providecommand{\bibinfo}[2]{#2}
\providecommand{\eprint}[2][]{\url{#2}}

\bibitem{Bazant2013}
\bibinfo{author}{Bazant, M.~Z.}
\newblock \bibinfo{title}{Theory of chemical kinetics and charge transfer based
  on nonequilibrium thermodynamics}.
\newblock \emph{\bibinfo{journal}{Acc. Chem. Res.}}
  \textbf{\bibinfo{volume}{46}}, \bibinfo{pages}{1144--1160}
  (\bibinfo{year}{2013}).

\bibitem{Latz2011}
\bibinfo{author}{Latz, A.} \& \bibinfo{author}{Zausch, J.}
\newblock \bibinfo{title}{{Thermodynamic consistent transport theory of Li-ion
  batteries}}.
\newblock \emph{\bibinfo{journal}{J. Power Sources}}
  \textbf{\bibinfo{volume}{196}}, \bibinfo{pages}{3296--3302}
  (\bibinfo{year}{2011}).

\bibitem{Hatzell2020}
\bibinfo{author}{Hatzell, K.~B.} \emph{et~al.}
\newblock \bibinfo{title}{{Challenges in Lithium Metal Anodes for Solid-State
  Batteries}}.
\newblock \emph{\bibinfo{journal}{ACS Energy Lett.}}
  \textbf{\bibinfo{volume}{5}}, \bibinfo{pages}{922--934}
  (\bibinfo{year}{2020}).

\bibitem{Cogswell2012}
\bibinfo{author}{Cogswell, D.~A.} \& \bibinfo{author}{Bazant, M.~Z.}
\newblock \bibinfo{title}{{Coherency Strain and the Kinetics of Phase
  Separation in LiFePO$_4$ Nanoparticles}}.
\newblock \emph{\bibinfo{journal}{ACS Nano}} \textbf{\bibinfo{volume}{6}},
  \bibinfo{pages}{2215--2225} (\bibinfo{year}{2012}).

\bibitem{Maier2004}
\bibinfo{author}{Maier, J.}
\newblock \emph{\bibinfo{title}{Physical Chemistry of Ionic Materials}}
  (\bibinfo{publisher}{John Wiley \& Sons, Ltd}, \bibinfo{year}{2004}).

\bibitem{Li2014}
\bibinfo{author}{Li, Y.} \emph{et~al.}
\newblock \bibinfo{title}{{Current-induced transition from particle-by-particle
  to concurrent intercalation in phase-separating battery electrodes}}.
\newblock \emph{\bibinfo{journal}{Nat. Mater.}} \textbf{\bibinfo{volume}{13}},
  \bibinfo{pages}{1149--1156} (\bibinfo{year}{2014}).

\bibitem{Liu2014}
\bibinfo{author}{Liu, H.} \emph{et~al.}
\newblock \bibinfo{title}{{Capturing metastable structures during high-rate
  cycling of LiFePO$_4$ nanoparticle electrodes}}.
\newblock \emph{\bibinfo{journal}{Science}} \textbf{\bibinfo{volume}{344}},
  \bibinfo{pages}{1252817} (\bibinfo{year}{2014}).

\bibitem{Lim2016}
\bibinfo{author}{Lim, J.} \emph{et~al.}
\newblock \bibinfo{title}{{Origin and hysteresis of lithium compositional
  spatiodynamics within battery primary particles}}.
\newblock \emph{\bibinfo{journal}{Science}} \textbf{\bibinfo{volume}{353}},
  \bibinfo{pages}{566--571} (\bibinfo{year}{2016}).

\bibitem{Gent2016}
\bibinfo{author}{Gent, W.~E.} \emph{et~al.}
\newblock \bibinfo{title}{{Persistent State-of-Charge Heterogeniety in Relaxed,
  Partially Charged Li$_{1-x}$Ni$_{1/3}$Co$_{1/3}$Mn$_{1/3}$O$_2$ Secondary
  Particles}}.
\newblock \emph{\bibinfo{journal}{Adv. Mater.}} \textbf{\bibinfo{volume}{28}},
  \bibinfo{pages}{6631--6638} (\bibinfo{year}{2016}).

\bibitem{Grenier2020}
\bibinfo{author}{Grenier, A.} \emph{et~al.}
\newblock \bibinfo{title}{{Intrinsic Kinetic Limitations in Substituted
  Lithium-Layered Transition-Metal Oxide Electrodes}}.
\newblock \emph{\bibinfo{journal}{J. Am. Chem. Soc.}}
  \textbf{\bibinfo{volume}{142}}, \bibinfo{pages}{7001--7011}
  (\bibinfo{year}{2020}).

\bibitem{Park2021}
\bibinfo{author}{Park, J.} \emph{et~al.}
\newblock \bibinfo{title}{{Fictitious phase separation in Li layered oxides
  driven by electro-autocatalysis}}.
\newblock \emph{\bibinfo{journal}{Nat. Mater.}} \textbf{\bibinfo{volume}{20}},
  \bibinfo{pages}{991--999} (\bibinfo{year}{2021}).

\bibitem{Xu2022}
\bibinfo{author}{Xu, C.} \emph{et~al.}
\newblock \bibinfo{title}{{Operando visualization of kinetically induced
  lithium heterogeneities in single-particle layered Ni-rich cathodes}}.
\newblock \emph{\bibinfo{journal}{Joule}} \textbf{\bibinfo{volume}{6}},
  \bibinfo{pages}{2535--2546} (\bibinfo{year}{2022}).

\bibitem{Huang2001}
\bibinfo{author}{Huang, H.}, \bibinfo{author}{Yin, S.-C.} \&
  \bibinfo{author}{Nazar, L.~F.}
\newblock \bibinfo{title}{{Approaching Theoretical Capacity of LiFePO$_4$ at
  Room Temperature at High Rates}}.
\newblock \emph{\bibinfo{journal}{Electrochem. Solid-State Lett.}}
  \textbf{\bibinfo{volume}{4}}, \bibinfo{pages}{A170} (\bibinfo{year}{2001}).

\bibitem{Li2017}
\bibinfo{author}{Li, J.} \emph{et~al.}
\newblock \bibinfo{title}{{Comparison of Single Crystal and Polycrystalline
  LiNi$_0.5$Mn$_0.3$Co$_0.2$O$_2$ Positive Electrode Materials for High Voltage
  Li-Ion Cells}}.
\newblock \emph{\bibinfo{journal}{J. Electrochem. Soc.}}
  \textbf{\bibinfo{volume}{164}}, \bibinfo{pages}{A1534}
  (\bibinfo{year}{2017}).

\bibitem{Lun2021}
\bibinfo{author}{Lun, Z.} \emph{et~al.}
\newblock \bibinfo{title}{{Cation-disordered rocksalt-type high-entropy
  cathodes for Li-ion batteries}}.
\newblock \emph{\bibinfo{journal}{Nat. Mater.}} \textbf{\bibinfo{volume}{20}},
  \bibinfo{pages}{214--221} (\bibinfo{year}{2021}).

\bibitem{Sada2023}
\bibinfo{author}{Sada, K.}, \bibinfo{author}{Darga, J.} \&
  \bibinfo{author}{Manthiram, A.}
\newblock \bibinfo{title}{{Challenges and Prospects of Sodium-Ion and
  Potassium-Ion Batteries for Mass Production}}.
\newblock \emph{\bibinfo{journal}{Adv. Energy Mater.}}
  \textbf{\bibinfo{volume}{13}}, \bibinfo{pages}{2302321}
  (\bibinfo{year}{2023}).

\bibitem{Dhir2020}
\bibinfo{author}{Dhir, S.}, \bibinfo{author}{Wheeler, S.},
  \bibinfo{author}{Capone, I.} \& \bibinfo{author}{Pasta, M.}
\newblock \bibinfo{title}{Outlook on {K}-ion batteries}.
\newblock \emph{\bibinfo{journal}{Chem}} \textbf{\bibinfo{volume}{6}},
  \bibinfo{pages}{2442--2460} (\bibinfo{year}{2020}).

\bibitem{IEA}
\bibinfo{author}{IEA}.
\newblock \bibinfo{title}{Global critical minerals outlook 2025}
  (\bibinfo{year}{2025}).
\newblock
  \urlprefix\url{{https://www.iea.org/reports/global-critical-minerals-outlook-2025}}.

\bibitem{Lu2012}
\bibinfo{author}{Lu, Y.}, \bibinfo{author}{Wang, L.}, \bibinfo{author}{Cheng,
  J.} \& \bibinfo{author}{Goodenough, J.~B.}
\newblock \bibinfo{title}{Prussian blue: a new framework of electrode materials
  for sodium batteries}.
\newblock \emph{\bibinfo{journal}{Chem. Commun.}}
  \textbf{\bibinfo{volume}{48}}, \bibinfo{pages}{6544--6546}
  (\bibinfo{year}{2012}).

\bibitem{Pasta2014}
\bibinfo{author}{Pasta, M.} \emph{et~al.}
\newblock \bibinfo{title}{{Full open-framework batteries for stationary energy
  storage}}.
\newblock \emph{\bibinfo{journal}{Nat. Commun.}} \textbf{\bibinfo{volume}{5}},
  \bibinfo{pages}{3007} (\bibinfo{year}{2014}).

\bibitem{Hurlbutt2018}
\bibinfo{author}{Hurlbutt, K.}, \bibinfo{author}{Wheeler, S.},
  \bibinfo{author}{Capone, I.} \& \bibinfo{author}{Pasta, M.}
\newblock \bibinfo{title}{{Prussian Blue Analogs as Battery Materials}}.
\newblock \emph{\bibinfo{journal}{Joule}} \textbf{\bibinfo{volume}{2}},
  \bibinfo{pages}{1950--1960} (\bibinfo{year}{2018}).

\bibitem{Wessells2011b}
\bibinfo{author}{Wessells, C.~D.}, \bibinfo{author}{Peddada, S.~V.},
  \bibinfo{author}{McDowell, M.~T.}, \bibinfo{author}{Huggins, R.~A.} \&
  \bibinfo{author}{Cui, Y.}
\newblock \bibinfo{title}{{The Effect of Insertion Species on Nanostructured
  Open Framework Hexacyanoferrate Battery Electrodes}}.
\newblock \emph{\bibinfo{journal}{J. Electrochem. Soc.}}
  \textbf{\bibinfo{volume}{159}}, \bibinfo{pages}{A98--A103}
  (\bibinfo{year}{2011}).

\bibitem{Li2017a}
\bibinfo{author}{Li, W.} \emph{et~al.}
\newblock \bibinfo{title}{Chemically diverse and multifunctional hybrid
  organic--inorganic perovskites}.
\newblock \emph{\bibinfo{journal}{Nat. Rev. Mater.}}
  \textbf{\bibinfo{volume}{2}}, \bibinfo{pages}{16099} (\bibinfo{year}{2017}).

\bibitem{Sharpe1976}
\bibinfo{author}{Sharpe, A.~G.}
\newblock \emph{\bibinfo{title}{{The chemistry of cyano complexes of the
  transition metals}}}.
\newblock Organometallic chemistry (\bibinfo{publisher}{Academic Press},
  \bibinfo{address}{London ; New York}, \bibinfo{year}{1976}).

\bibitem{Wu2016}
\bibinfo{author}{Wu, X.} \emph{et~al.}
\newblock \bibinfo{title}{{Highly Crystallized Na$_2$CoFe(CN)$_6$ with
  Suppressed Lattice Defects as Superior Cathode Material for Sodium-Ion
  Batteries}}.
\newblock \emph{\bibinfo{journal}{ACS Appl. Mater. Interfaces}}
  \textbf{\bibinfo{volume}{8}}, \bibinfo{pages}{5393--5399}
  (\bibinfo{year}{2016}).

\bibitem{Moritomo2011}
\bibinfo{author}{Moritomo, Y.}, \bibinfo{author}{Kurihara, Y.},
  \bibinfo{author}{Matsuda, T.} \& \bibinfo{author}{Kim, J.}
\newblock \bibinfo{title}{{Structural phase diagram of Mn-Fe cyanide against
  cation concentration}}.
\newblock \emph{\bibinfo{journal}{J. Phys. Soc. Jpn.}}
  \textbf{\bibinfo{volume}{80}}, \bibinfo{pages}{103601}
  (\bibinfo{year}{2011}).

\bibitem{Cattermull2021}
\bibinfo{author}{Cattermull, J.}, \bibinfo{author}{Pasta, M.} \&
  \bibinfo{author}{Goodwin, A.~L.}
\newblock \bibinfo{title}{{Structural complexity in Prussian blue analogues}}.
\newblock \emph{\bibinfo{journal}{Mater. Horiz.}} \textbf{\bibinfo{volume}{8}},
  \bibinfo{pages}{3178--3186} (\bibinfo{year}{2021}).

\bibitem{Jiang2019}
\bibinfo{author}{Jiang, L.} \emph{et~al.}
\newblock \bibinfo{title}{{Building aqueous K-ion batteries for energy
  storage}}.
\newblock \emph{\bibinfo{journal}{Nat. Energy}} \textbf{\bibinfo{volume}{4}},
  \bibinfo{pages}{495--503} (\bibinfo{year}{2019}).

\bibitem{Cattermull2022}
\bibinfo{author}{Cattermull, J.} \emph{et~al.}
\newblock \bibinfo{title}{{Uncovering the Interplay of Competing Distortions in
  the Prussian Blue Analogue K$_2$Cu[Fe(CN)$_6$]}}.
\newblock \emph{\bibinfo{journal}{Chem. Mater.}} \textbf{\bibinfo{volume}{34}},
  \bibinfo{pages}{5000--5008} (\bibinfo{year}{2022}).

\bibitem{Cattermull2023a}
\bibinfo{author}{Cattermull, J.}, \bibinfo{author}{Pasta, M.} \&
  \bibinfo{author}{Goodwin, A.~L.}
\newblock \bibinfo{title}{{Predicting Distortion Magnitudes in Prussian Blue
  Analogues}}.
\newblock \emph{\bibinfo{journal}{J. Am. Chem. Soc.}}
  \textbf{\bibinfo{volume}{145}}, \bibinfo{pages}{24471--24475}
  (\bibinfo{year}{2023}).

\bibitem{Bie2017}
\bibinfo{author}{Bie, X.}, \bibinfo{author}{Kubota, K.},
  \bibinfo{author}{Hosaka, T.}, \bibinfo{author}{Chihara, K.} \&
  \bibinfo{author}{Komaba, S.}
\newblock \bibinfo{title}{{A novel K-ion battery: hexacyanoferrate(II)/graphite
  cell}}.
\newblock \emph{\bibinfo{journal}{J. Mater. Chem. A}}
  \textbf{\bibinfo{volume}{5}}, \bibinfo{pages}{4325--4330}
  (\bibinfo{year}{2017}).

\bibitem{Fiore2020}
\bibinfo{author}{Fiore, M.} \emph{et~al.}
\newblock \bibinfo{title}{{Paving the Way toward Highly Efficient, High-Energy
  Potassium-Ion Batteries with Ionic Liquid Electrolytes}}.
\newblock \emph{\bibinfo{journal}{Chem. Mater.}} \textbf{\bibinfo{volume}{32}},
  \bibinfo{pages}{7653--7661} (\bibinfo{year}{2020}).

\bibitem{Cattermull2023}
\bibinfo{author}{Cattermull, J.}, \bibinfo{author}{Roth, N.},
  \bibinfo{author}{Cassidy, S.~J.}, \bibinfo{author}{Pasta, M.} \&
  \bibinfo{author}{Goodwin, A.~L.}
\newblock \bibinfo{title}{{K-ion Slides in Prussian Blue Analogues}}.
\newblock \emph{\bibinfo{journal}{J. Am. Chem. Soc.}}
  \textbf{\bibinfo{volume}{145}}, \bibinfo{pages}{24249--24259}
  (\bibinfo{year}{2023}).

\bibitem{Wessells2011}
\bibinfo{author}{Wessells, C.~D.}, \bibinfo{author}{Huggins, R.~A.} \&
  \bibinfo{author}{Cui, Y.}
\newblock \bibinfo{title}{{Copper hexacyanoferrate battery electrodes with long
  cycle life and high power}}.
\newblock \emph{\bibinfo{journal}{Nat. Commun.}} \textbf{\bibinfo{volume}{2}},
  \bibinfo{pages}{550} (\bibinfo{year}{2011}).

\bibitem{Hosaka2021}
\bibinfo{author}{Hosaka, T.}, \bibinfo{author}{Fukabori, T.},
  \bibinfo{author}{Kojima, H.}, \bibinfo{author}{Kubota, K.} \&
  \bibinfo{author}{Komaba, S.}
\newblock \bibinfo{title}{Effect of particle size and anion vacancy on
  electrochemical potassium ion insertion into potassium manganese
  hexacyanoferrates}.
\newblock \emph{\bibinfo{journal}{ChemSusChem}} \textbf{\bibinfo{volume}{14}},
  \bibinfo{pages}{1166--1175} (\bibinfo{year}{2021}).

\bibitem{Deng2021}
\bibinfo{author}{Deng, L.} \emph{et~al.}
\newblock \bibinfo{title}{{Defect-free potassium manganese hexacyanoferrate
  cathode material for high-performance potassium-ion batteries}}.
\newblock \emph{\bibinfo{journal}{Nat. Commun.}} \textbf{\bibinfo{volume}{12}},
  \bibinfo{pages}{2167} (\bibinfo{year}{2021}).

\bibitem{Dhir2024}
\bibinfo{author}{Dhir, S.} \emph{et~al.}
\newblock \bibinfo{title}{{Characterisation and Modelling of Potassium-ion
  Batteries}}.
\newblock \emph{\bibinfo{journal}{Nat. Commun.}} \textbf{\bibinfo{volume}{15}},
  \bibinfo{pages}{7580} (\bibinfo{year}{2024}).

\bibitem{Wang2015}
\bibinfo{author}{Wang, L.} \emph{et~al.}
\newblock \bibinfo{title}{{Rhombohedral Prussian white as cathode for
  rechargeable sodium-ion batteries}}.
\newblock \emph{\bibinfo{journal}{J. Am. Chem. Soc.}}
  \textbf{\bibinfo{volume}{137}}, \bibinfo{pages}{2548--2554}
  (\bibinfo{year}{2015}).

\bibitem{Brant2019}
\bibinfo{author}{Brant, W.~R.} \emph{et~al.}
\newblock \bibinfo{title}{{Selective Control of Composition in Prussian White
  for Enhanced Material Properties}}.
\newblock \emph{\bibinfo{journal}{Chem. Mater.}} \textbf{\bibinfo{volume}{31}},
  \bibinfo{pages}{7203--7211} (\bibinfo{year}{2019}).

\bibitem{Jiang2017}
\bibinfo{author}{Jiang, X.}, \bibinfo{author}{Zhang, T.},
  \bibinfo{author}{Yang, L.}, \bibinfo{author}{Li, G.} \& \bibinfo{author}{Lee,
  J.~Y.}
\newblock \bibinfo{title}{{A Fe/Mn-Based Prussian Blue Analogue as a K-Rich
  Cathode Material for Potassium-Ion Batteries}}.
\newblock \emph{\bibinfo{journal}{ChemElectroChem}}
  \textbf{\bibinfo{volume}{4}}, \bibinfo{pages}{2237--2242}
  (\bibinfo{year}{2017}).

\bibitem{Harbourne2024a}
\bibinfo{author}{Harbourne, E.~A.} \emph{et~al.}
\newblock \bibinfo{title}{{Rules governing Jahn-Teller order in Prussian blue
  analogues}}.
\newblock \emph{\bibinfo{journal}{arXiv}} \bibinfo{pages}{DOI:
  10.48550/arXiv.2408.13169} (\bibinfo{year}{2024}).

\bibitem{Bostrom2022}
\bibinfo{author}{Bostr{\"{o}}m, H. L.~B.} \& \bibinfo{author}{Brant, W.~R.}
\newblock \bibinfo{title}{{Octahedral tilting in Prussian blue analogues}}.
\newblock \emph{\bibinfo{journal}{J. Mater. Chem. C}}
  \textbf{\bibinfo{volume}{10}}, \bibinfo{pages}{13690--13699}
  (\bibinfo{year}{2022}).

\bibitem{Peng2019}
\bibinfo{author}{Peng, F.} \emph{et~al.}
\newblock \bibinfo{title}{Highly crystalline sodium manganese ferrocyanide
  microcubes for advanced sodium ion battery cathodes}.
\newblock \emph{\bibinfo{journal}{J. Mater. Chem. A}}
  \textbf{\bibinfo{volume}{7}}, \bibinfo{pages}{22248--22256}
  (\bibinfo{year}{2019}).

\bibitem{LePham2023}
\bibinfo{author}{Le~Pham, P.~N.} \emph{et~al.}
\newblock \bibinfo{title}{Prussian blue analogues for potassium-ion batteries:
  insights into the electrochemical mechanisms}.
\newblock \emph{\bibinfo{journal}{J. Mater. Chem. A}}
  \textbf{\bibinfo{volume}{11}}, \bibinfo{pages}{3091--3104}
  (\bibinfo{year}{2023}).

\bibitem{Geddes2019}
\bibinfo{author}{Geddes, H.~S.}, \bibinfo{author}{Blade, H.},
  \bibinfo{author}{McCabe, J.~F.}, \bibinfo{author}{Hughes, L.~P.} \&
  \bibinfo{author}{Goodwin, A.~L.}
\newblock \bibinfo{title}{Structural characterisation of amorphous solid
  dispersions via metropolis matrix factorisation of pair distribution function
  data}.
\newblock \emph{\bibinfo{journal}{Chem. Commun.}}
  \textbf{\bibinfo{volume}{55}}, \bibinfo{pages}{13346--13349}
  (\bibinfo{year}{2019}).

\bibitem{Padhi1997}
\bibinfo{author}{Padhi, A.~K.}, \bibinfo{author}{Nanjundaswamy, K.~S.} \&
  \bibinfo{author}{Goodenough, J.~B.}
\newblock \bibinfo{title}{{Phospho‐olivines as Positive‐Electrode Materials
  for Rechargeable Lithium Batteries}}.
\newblock \emph{\bibinfo{journal}{J. Electrochem. Soc.}}
  \textbf{\bibinfo{volume}{144}}, \bibinfo{pages}{1188} (\bibinfo{year}{1997}).

\bibitem{Harbourne2024}
\bibinfo{author}{Harbourne, E.~A.} \emph{et~al.}
\newblock \bibinfo{title}{{Local structure and dynamics in MPt(CN)$_6$ Prussian
  blue analogues}}.
\newblock \emph{\bibinfo{journal}{Chem. Mater.}} \textbf{\bibinfo{volume}{36}},
  \bibinfo{pages}{5796--5804} (\bibinfo{year}{2024}).

\bibitem{Xu2017}
\bibinfo{author}{Xu, G.-L.} \emph{et~al.}
\newblock \bibinfo{title}{Insights into the structural effects of layered
  cathode materials for high voltage sodium-ion batteries}.
\newblock \emph{\bibinfo{journal}{Energy Environ. Sci.}}
  \textbf{\bibinfo{volume}{10}}, \bibinfo{pages}{1677--1693}
  (\bibinfo{year}{2017}).
\newblock \urlprefix\url{http://dx.doi.org/10.1039/C7EE00827A}.

\bibitem{Bruce2008}
\bibinfo{author}{Bruce, P.}, \bibinfo{author}{Scrosati, B.} \&
  \bibinfo{author}{Tarascon, J.-M.}
\newblock \bibinfo{title}{Nanomaterials for rechargeable lithium batteries}.
\newblock \emph{\bibinfo{journal}{Angew. Chem. Int. Ed.}}
  \textbf{\bibinfo{volume}{47}}, \bibinfo{pages}{2930--2946}
  (\bibinfo{year}{2008}).

\bibitem{Simonov2020}
\bibinfo{author}{Simonov, A.} \emph{et~al.}
\newblock \bibinfo{title}{Hidden diversity of vacancy networks in {P}russian
  blue analogues}.
\newblock \emph{\bibinfo{journal}{Nature}} \textbf{\bibinfo{volume}{578}},
  \bibinfo{pages}{256--260} (\bibinfo{year}{2020}).

\bibitem{Tan2011}
\bibinfo{author}{Tan, J.~C.} \& \bibinfo{author}{Cheetham, A.~K.}
\newblock \bibinfo{title}{Mechanical properties of hybrid inorganic--organic
  framework materials: establishing fundamental structure--property
  relationships}.
\newblock \emph{\bibinfo{journal}{Chem. Soc. Rev.}}
  \textbf{\bibinfo{volume}{40}}, \bibinfo{pages}{1059--1080}
  (\bibinfo{year}{2011}).

\bibitem{Liu2008}
\bibinfo{author}{Liu, Y.} \emph{et~al.}
\newblock \bibinfo{title}{{Reversible Structural Transition in MIL-53 with
  Large Temperature Hysteresis}}.
\newblock \emph{\bibinfo{journal}{J. Am. Chem. Soc.}}
  \textbf{\bibinfo{volume}{130}}, \bibinfo{pages}{11813--11818}
  (\bibinfo{year}{2008}).

\bibitem{Krause2016}
\bibinfo{author}{Krause, S.} \emph{et~al.}
\newblock \bibinfo{title}{A pressure-amplifying framework material with
  negative gas adsorption transitions}.
\newblock \emph{\bibinfo{journal}{Nature}} \textbf{\bibinfo{volume}{532}},
  \bibinfo{pages}{348--352} (\bibinfo{year}{2016}).

\bibitem{Guzelturk2021}
\bibinfo{author}{Guzelturk, B.} \emph{et~al.}
\newblock \bibinfo{title}{Visualization of dynamic polaronic strain fields in
  hybrid lead halide perovskites}.
\newblock \emph{\bibinfo{journal}{Nat. Mater.}} \textbf{\bibinfo{volume}{20}},
  \bibinfo{pages}{618--623} (\bibinfo{year}{2021}).
\newblock \urlprefix\url{https://doi.org/10.1038/s41563-020-00865-5}.

\bibitem{Dubajic2025}
\bibinfo{author}{Dubajic, M.} \emph{et~al.}
\newblock \bibinfo{title}{Dynamic nanodomains dictate macroscopic properties in
  lead halide perovskites}.
\newblock \emph{\bibinfo{journal}{Nat. Nanotechnol.}}
  \textbf{\bibinfo{volume}{6}}, \bibinfo{pages}{755--763}
  (\bibinfo{year}{2025}).

\bibitem{Saparov2016}
\bibinfo{author}{Saparov, B.} \& \bibinfo{author}{Mitzi, D.~B.}
\newblock \bibinfo{title}{Organic--inorganic perovskites: Structural
  versatility for functional materials design}.
\newblock \emph{\bibinfo{journal}{Chem. Rev.}} \textbf{\bibinfo{volume}{116}},
  \bibinfo{pages}{4558--4596} (\bibinfo{year}{2016}).

\bibitem{Bennett2017}
\bibinfo{author}{D., B.~T.}, \bibinfo{author}{Cheeetham, A.~K.},
  \bibinfo{author}{Fuchs, A.~H.} \& \bibinfo{author}{Coudert, F.-X.}
\newblock \bibinfo{title}{Interplay between defects, disorder and flexibility
  in metal-organic frameworks}.
\newblock \emph{\bibinfo{journal}{Nat. Chem.}} \textbf{\bibinfo{volume}{9}},
  \bibinfo{pages}{11--16} (\bibinfo{year}{2017}).

\bibitem{Bostrom2018}
\bibinfo{author}{Bostr{\"{o}}m, H. L.~B.}, \bibinfo{author}{Senn, M.~S.} \&
  \bibinfo{author}{Goodwin, A.~L.}
\newblock \bibinfo{title}{Recipes for improper ferroelectricity in molecular
  perovskites}.
\newblock \emph{\bibinfo{journal}{Nat. Commun.}} \textbf{\bibinfo{volume}{9}},
  \bibinfo{pages}{2380} (\bibinfo{year}{2018}).

\bibitem{Kronawitter2024}
\bibinfo{author}{Kronawiiter, S.~M.} \& \bibinfo{author}{Kieslich, G.}
\newblock \bibinfo{title}{The wondrous world of {ABX}$_3$ molecular
  perovskites}.
\newblock \emph{\bibinfo{journal}{Chem. Commun.}}
  \textbf{\bibinfo{volume}{60}}, \bibinfo{pages}{11673--11684}
  (\bibinfo{year}{2024}).

\end{thebibliography}

\end{document}